\documentclass[aps,pre,reprint,superscriptaddress,longbibliography,floatfix]{revtex4-2}
\usepackage{amsmath,amssymb}
\usepackage{graphicx}
\usepackage{algorithm,algpseudocode}
\usepackage{xcolor}
\usepackage[colorlinks=true,citecolor=blue,linkcolor=blue,urlcolor=blue]{hyperref}

\begin{document}

\title{Critical geometry of triconnected clusters in percolation}

\date{\today}

\author{Qi Wang}
\affiliation{School of Physics, Hefei University of Technology, Hefei, Anhui 230009, China}

\author{Youjin Deng}
\email{yjdeng@ustc.edu.cn}
\affiliation{Hefei National Research Center for Physical Sciences at the Microscale, University of Science and Technology of China, Hefei 230026, China}
\affiliation{Department of Modern Physics, University of Science and Technology of China, Hefei 230026, China}
\affiliation{Hefei National Laboratory, University of Science and Technology of China, Hefei 230088, China}

\author{Ming Li}
\email{lim@hfut.edu.cn}
\affiliation{School of Physics, Hefei University of Technology, Hefei, Anhui 230009, China}

\begin{abstract}
We investigate the critical geometry of triconnected clusters in percolation, which remain connected after the removal of any two sites. By combining biconnected-component decomposition with the SPQR decomposition of graphs, we develop an efficient algorithm for identifying triconnected clusters in large systems. Extensive Monte Carlo simulations yield estimates of the fractal dimension $d_3=1.2308(3)$, $1.083(2)$, and $0.77(2)$ in dimensions $d=2$, $3$, and $4$, respectively, revealing a systematic decrease of $d_3$ with increasing $d$. In dimension $d=6$, the largest triconnected cluster grows logarithmically with system size, whereas on complete graphs, corresponding to $d=\infty$, it approaches a finite size. Both behaviors indicate a vanishing fractal dimension, $d_3=0$, for $d\geq6$. Furthermore, the singular contribution to the susceptibility-like quantity associated with the second moment of the triconnected-cluster size distribution scales as $\chi\sim L^{2d_3-d}$. Consequently, it vanishes for $d\geq3$, leaving a finite asymptotic susceptibility, in contrast to the divergent behavior observed in standard percolation.
 \end{abstract}

\maketitle

\section{Introduction}

Phase transitions constitute a central framework for understanding collective behavior in statistical physics. Beyond thermodynamic quantities and conventional order parameters, phase transitions are often accompanied by characteristic changes in the geometry of microscopic structures. Percolation theory~\cite{Stauffer1994}, which considers the connected clusters formed by occupied sites or bonds, provides a prototypical setting for studying these geometric aspects of critical phenomena. At criticality, these clusters exhibit scale-invariant structures characterized by nontrivial fractal dimensions. Importantly, such geometric structures are not restricted to percolation models themselves. Through the Fortuin-Kasteleyn (FK) representation~\cite{Kasteleyn1969,Fortuin1972}, analogous cluster structures can be constructed in a broad class of spin models, establishing a direct connection between cluster geometry and thermodynamic critical behavior. However, the size and fractal dimension of a cluster characterize its overall geometry but do not fully resolve its internal connectivity~\cite{Gyure1995,Aizenman1999,Xu2014}.

\begin{figure}
\centering
\includegraphics[width=\columnwidth]{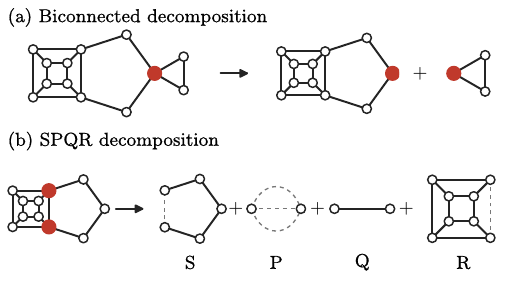}
\caption{Sketch of connectivity decomposition. (a) An ordinary connected cluster is decomposed at the red articulation points into two biconnected components, with each articulation point retained in both components. (b) A biconnected component is decomposed at the red separation pair. The two sites are retained as poles in each resulting subgraph and connected by a virtual bond (dashed line), representing the remainder of the original component. The resulting subgraphs are single bonds (Q), cycles (S), R-components, or P-components. An R-component is a triconnected cluster without separation pairs, whereas a P-component consists of parallel bonds between the two poles. Although the original graph is simple, virtual bonds can generate parallel bonds and hence P-components.} \label{f1}
\end{figure}

A natural way to characterize the internal geometry of a cluster is through the robustness of its connectivity. More generally, a cluster is said to be $k$-connected if it remains connected after the removal of any $k-1$ sites~\cite{Tutte2001}. For example, as illustrated in Fig.~\ref{f1}(a), removing the red site, known as an articulation point (or cut vertex), disconnects the cluster. Decomposing the cluster at all its articulation points yields biconnected components, which contain no single-site bottlenecks. However, as shown in Fig.~\ref{f1}(b), a biconnected component can still be disconnected by removing a pair of sites (red sites), known as a separation pair. Recursively decomposing the biconnected component at such separation pairs gives rise to triconnected components. This defines a hierarchy of connectivity: biconnected structures are free of single-site bottlenecks and are closely related to the backbone~\cite{HerrmannStanley1984,Deng2004a,Fang2022}, while triconnected structures are free of two-site bottlenecks. Higher connectivity therefore provides a systematic way to resolve increasingly robust structures embedded within a critical cluster.

In two dimensions, conformal field theory (CFT) and Schramm–Loewner evolution (SLE) have provided exact results for many critical exponents associated with conventional connectivity~\cite{Nienhuis1982,Smirnov2001,Lawler2002}. In contrast, higher-order connected structures remain poorly understood, particularly for large $k$. Their geometry can be characterized through the monochromatic $k$-arm exponents, $\xi_k$, which describe the probability of finding $k$ disjoint paths of the same color extending from a given site to a distance $r$, with a power-law decay $r^{-\xi_k}$~\cite{Berg2020,Beffara2011}. The corresponding fractal dimension $d_k$ of $k$-connected component is related to the arm exponent through $d_k=d-\xi_k$ in $d$ dimensions. A striking example is the backbone ($k=2$), whose fractal dimension was known only numerically for decades~\cite{HerrmannStanley1984,Jacobsen2002,Deng2004a,Fang2022,Xu2014}, before being derived exactly and shown to be a transcendental number~\cite{Nolin2023,Nolin2025}. This result is particularly notable because many previously known exactly determined critical exponents were rational numbers. It highlights the possibility that stronger connectivity constraints give rise to nontrivial geometric properties that are not readily captured by the conventional description of critical clusters. This raises a fundamental physical question: how does higher-order connectivity reshape the geometry and scaling behavior of critical clusters?

For triconnected clusters and high-order connectivities ($k\geq3$), corresponding questions remain largely unexplored. We approach this problem through Monte Carlo simulations and finite-size scaling (FSS), aiming to establish whether triconnected clusters exhibit well-defined fractal scaling and to characterize how their fractal dimension depends on spatial dimension. Such a quantitative characterization provides a basis for comparing different levels of connectivity and understanding how increasingly stringent connectivity constraints affect critical geometry.

Ordinary connected clusters can be identified using breadth-first search or depth-first search (DFS) in $\mathcal{O}(V)$ time, where $V$ is the number of sites. Biconnected structures can likewise be extracted in linear time using modified DFS algorithms~\cite{Tarjan1972,Fang2022}. In contrast, identifying triconnected clusters requires more sophisticated graph algorithms to locate separation pairs, typically through SPQR decomposition~\cite{Hopcroft1973,DiBattista1996,Gutwenger2001,Mutzel2003}.

The additional computational cost has limited large-scale numerical studies, and previous estimates of the triconnected-cluster fractal dimension in two dimensions have consequently remained relatively imprecise, with reported values of $d_3=1.2(1)$~\cite{Paul2002}, $1.23(2)$~\cite{Jacobsen2002}, and $1.248(8)$~\cite{MohseniKabir2021}. Research on high-dimensional cases is even scarcer, while fractal dimensions remain only roughly estimated~\cite{Paul2003}.

\begin{table}
\caption{Estimated fractal dimensions $d_3$ of triconnected clusters in dimensions $d=2$, $3$ and $4$. Results obtained from site percolation on triangular lattices for $d=2$, and from bond percolation on cubic or hypercubic lattices for $d=3$ and $4$. For comparison, the previous estimates of $d_3$, as well as the fractal dimensions (numerical estimates or exact solutions) of connected and biconnected clusters, $d_1$ and $d_2$, are also listed.}     \label{tab-all}
\begin{ruledtabular}
\begin{tabular}{clll}
$d$   & \multicolumn{1}{c}{$d_3$}  & \multicolumn{1}{c}{$d_2$}  & \multicolumn{1}{c}{$d_1$}  \\
\hline
$2$      & $1.2(1)$~\cite{Paul2002}              & $1.64336(10)$~\cite{Xu2014}  & $91/48$~\cite{Coniglio1989,Smirnov2001}    \\
         & $1.23(2)$~\cite{Jacobsen2002}         & $1.64339(5)$~\cite{Fang2022}       &     \\
         & $1.248(8)$~\cite{MohseniKabir2021}    & $1.64333\ldots$~\cite{Nolin2025}      &     \\
present  & $1.2308(3)$    &     &     \\
\hline
$3$      & $1.15(10)$~\cite{Paul2002}            & $1.87(3)$~\cite{Porto1997}       & $2.5230(1)$~\cite{Ballesteros1999}   \\
         &                                       & $1.875(3)$~\cite{Deng2004}       & $2.5226(1)$~\cite{Deng2005}     \\
         &                                       & $1.855(15)$~\cite{Rintoul1994}   & $2.52293(10)$~\cite{Xu2013}     \\
present  & $1.083(2)$    &     &     \\
\hline
$4$      & $0.7(2)$~\cite{Paul2003}    &  $1.9844(11)$~\cite{Zhang2021}     &  $3.046(7)$\cite{Paul2001}  \\
         &                             &                                    &  $3.0437(11)$\cite{Mertens2018}   \\
         &                             &                                    &  $3.0446(7)$~\cite{Zhang2021}  \\
present  & $0.77(2)$    &     &
\end{tabular}
\end{ruledtabular}
\end{table}

In this paper, we perform large-scale numerical studies of triconnected clusters in percolation in dimensions $d=2$, $3$, $4$, $6$, and on complete graphs. To make these calculations feasible, we implement an SPQR-based decomposition algorithm that identifies triconnected clusters in approximately linear time. We determine their fractal dimensions with high precision, as list in Table~\ref{tab-all}, respectively. In dimension $d=6$, we find that the triconnected cluster grows logarithmically with linear size $L$, corresponding to a vanishing power-law fractal dimension. On complete graphs, its size instead approaches a finite value as the system size increases. Remarkably, these changes in the scaling of triconnected clusters occur even though the percolation transition remains continuous. For $d<6$, the size distributions are consistent with the standard scaling form, with Fisher exponents satisfying the hyperscaling relation $\tau_3=1+d/d_3$. Small $d_3$ further predicts a second moment whose singular contribution grows as $L^{2d_3-d}$ in $d=2$ but vanishes for $d\geq 3$, leaving a finite background. Together, these results provide a quantitative picture of the critical geometry of triconnected clusters and high-precision numerical benchmarks for higher-connectivity critical structures.

The paper is organized as follows. In Sec.~\ref{sec-ma}, we introduce the percolation model and the algorithm to identify triconnected clusters in critical percolation configurations. The simulation results and FSS analysis are given in Sec.~\ref{sec-sr}. Finally, Sec.~\ref{sec-dis} presents a summary and a discussion.

\section{Model and algorithm}   \label{sec-ma}

\subsection{Percolation systems}

We study triconnected clusters in critical percolation configurations in dimensions $d=2$, $3$, $4$, and $6$, as well as on complete graphs. The side length of lattices in dimension $d$ denotes as $L$, and thus the total number of sites are $V=L^d$. In site (bond) percolation, each site (bond) is independently occupied with probability $p$. We focus on the configuration at percolation threshold $p_c$. In two dimensions, we consider bond percolation on the square lattice and site percolation on the triangular lattice, both of which have a threshold $p_c=1/2$~\cite{Stauffer1994}. In dimensions $d=3$, $4$, and $6$, we consider bond percolation on hypercubic lattices, with thresholds $p_c=0.24881182$~\cite{Wang2013}, $0.16013122$~\cite{Mertens2018,Xun2020}, and $0.09420165$~\cite{Mertens2018}, respectively. For complete graphs, the critical point is given by $p_c=1/(V-1)$, corresponding to a critical average degree $p_c(V-1)=1$.

\subsection{Algorithm}

For each percolation configuration, we first identify the biconnected components of each ordinary connected cluster. This can be achieved in linear time using DFS by identifying all articulation points, whose removal disconnects a cluster~\cite{Tarjan1972}; see Fig.~\ref{f1}(a). During the DFS, each site $u$ is assigned a discovery time $\operatorname{dfn}[u]$. We also record $\operatorname{low1}[u]$, the smallest discovery time of an ancestor reachable from the subtree rooted at $u$ by following tree bonds and then using at most one back bond, excluding the parent tree bond. If no such ancestor is reached, we set $\operatorname{low1}[u]=\operatorname{dfn}[u]$. A non-root site $u$ is an articulation point if at least one of its DFS-tree children $v$ satisfies
\begin{equation}
\operatorname{low1}[v]\geq \operatorname{dfn}[u].
\end{equation}
The root is an articulation point if it has at least two DFS-tree children. These conditions allow all articulation points and hence all biconnected components to be identified in linear time.

\begin{figure}
\centering
\includegraphics[width=\columnwidth]{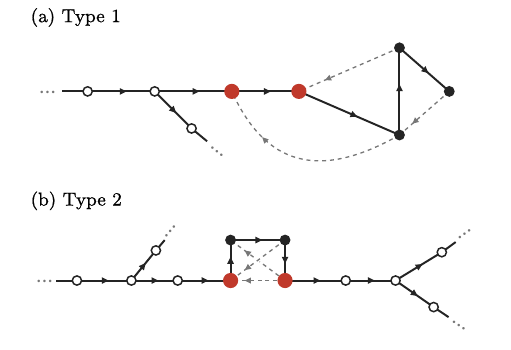}
\caption{Two types of separation pairs in a DFS tree. The solid lines with arrows represent the paths of DFS, while the dashed lines represent the backward bonds. The red sites indicate the two sites forming a separation pair. (a) Type 1: the separation pair separates a subtree of the DFS tree (black solid points, along with the separation pair). (b) Type 2: the separation pair separates a set of sites (black solid points, along with the separation pair) lying between the two sites along a path in the DFS tree.} \label{f2}
\end{figure}

\begin{figure}
\centering
\includegraphics[width=\columnwidth]{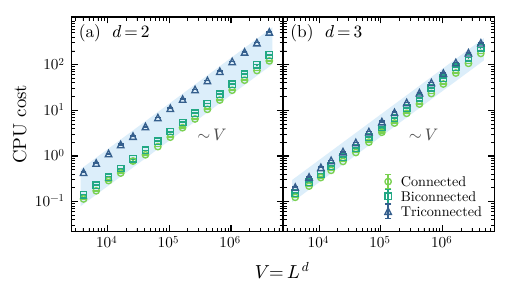}
\caption{CPU cost required to identify connected, biconnected, and triconnected clusters in (a) critical site percolation on the triangular lattice and (b) critical bond percolation on the cubic lattice. Here, $V=L^d$ denotes the total number of lattice sites for a system of linear size $L$. Each point is the mean computation time (in milliseconds) over 12 independent percolation configurations. Measurements use a single thread on an Intel Core i7-12700K processor.}   \label{f3}
\end{figure}

We then decompose each biconnected component into triconnected components using an SPQR-based algorithm~\cite{Hopcroft1973,DiBattista1996,Gutwenger2001,Mutzel2003}. The algorithm first identifies separation pairs, whose removal disconnects a biconnected component, which can be also realized by DFS. In addition to $\operatorname{dfn}[u]$ and $\operatorname{low1}[u]$, we record $\operatorname{low2}[u]$, the second-smallest distinct discovery time of an ancestor reachable from the subtree rooted at $u$ through tree bonds followed by at most one back bond. If fewer than two distinct earlier targets are available, the missing value is set to $\operatorname{dfn}[u]$. These two values provide the information needed to identify separation pairs. Specifically, consider a DFS-tree bond from a parent site $v$ to its child $u$, if
\begin{equation}
\operatorname{low1}[u] < \operatorname{dfn}[v],
\qquad
\operatorname{low2}[u] \geq \operatorname{dfn}[v],
\end{equation}
then the ancestor $w$ satisfying $\operatorname{dfn}[w]=\operatorname{low1}[u]$ and site $v$ form a separation pair, provided that at least one site remains on each side after removing $v$ and $w$.

Note that this algorithm does not identify all separation pairs, but only those required to decompose the graph into its split components. In general, two types of such separation pairs can be identified in the DFS tree, as illustrated in Fig.~\ref{f2}. A type-1 separation pair separates the subtree rooted at $u$ from the rest of the component, whereas a type-2 separation pair separates a subgraph associated with the path between $w$ and $v$ in the DFS tree. Further details of the algorithm can be found in Refs.~\cite{Hopcroft1973,Gutwenger2001,Chimani2014OGDF}.

When a separation pair is identified, the two sites are retained in each resulting subgraph as poles and are connected by a virtual bond [dashed lines in Fig.~\ref{f1}(b)]. This virtual bond represents the part of the original component lying outside the corresponding subgraph. The decomposition proceeds until the resulting skeletons are classified as cycles (S), parallel bonds (P), single bonds (Q), or triconnected clusters (R). It also requires a separation pair to split a component into nontrivial subgraphs; thus, an individual bond is not separated from an R-component or an S-component merely because its two endpoints can be chosen as a pair of sites.

Overall, both biconnected and triconnected clusters can be identified via DFS, and the associated computational cost scales approximately linearly with system size. Figure~\ref{f3} shows the CPU cost, using $d=2$ and $3$ as examples. For biconnected clusters, the computational cost is only marginally higher than for ordinary connected clusters, as the additional DFS-related operations are relatively few. For triconnected cluster, it takes approximately four times as long as ordinary cluster identification in two dimensions, whereas in three dimensions the ratio is about two. This difference is partly related to the structure of percolation clusters: for a fixed total number of lattice sites, higher-dimensional systems contain substantially smaller biconnected and triconnected components, as shown below, reducing the computational effort required for their decomposition. Nevertheless, the measured CPU cost increase approximately linearly with the total number of lattice sites $V$ in both dimensions, demonstrating that the algorithm is efficient for large-scale simulations.

\section{Observables and fitting protocol}

For each configuration of triconnected clusters, we measure the following quantities:
\begin{itemize}
  \item Let $\mathcal{C}_1$ denote the size of the largest triconnected cluster in a given realization, counting the number of bonds. The ensemble-averaged size is defined as
      \begin{equation}
      C_1 \equiv \langle \mathcal{C}_1 \rangle.
      \end{equation}

  \item The probability density $P(x)$ of the rescaled $C_1$,
      \begin{equation}
      x \equiv \frac{C_1}{L^{d_3}},
      \end{equation}
        where $d_3$ is the fractal dimension of triconnected clusters.

  \item The susceptibility-like quantity,
      \begin{equation}
      \chi\equiv\frac{1}{V}\left\langle\sum_{i\ne1}\mathcal{C}_i^2\right\rangle,
      \label{eq_order_chi}
      \end{equation}
      where the sum runs over all cluster sizes $\mathcal{C}_i$ with the largest one excluded.

  \item The size distribution of triconnected clusters,
      \begin{equation}
      n(s,L) = \frac{\langle \mathcal{N}_s(\Delta s) \rangle}{V \, \Delta s},
      \end{equation}
      where $\mathcal{N}_s(\Delta s)$ is the number of triconnected clusters with sizes in the interval $[s, s+\Delta s)$. We use geometrical binning with $\Delta s = a^k$ for the $k$-th bin starting from $s=1$, where $a>1$ controls the binning resolution. The bin center is defined as $s'=\sqrt{s(s+\Delta s)}$.
\end{itemize}
Here, $\langle \cdot \rangle$ denotes averaging over independent realizations. The simulation results are obtained by averaging over realizations varying from $10^5$ to $10^8$ times. Owing to time complexity, larger systems are constrained to use only a small number of averaging realizations, while smaller systems can average over many more realizations.

To estimate critical exponents, such as $d_3$, we fit the data to the FSS form
\begin{equation}
Q(L)=L^{X_Q}\left(a_0+a_1L^{-\omega}\right), \label{eq-fit}
\end{equation}
where $Q$ denotes the observable and $X_Q$ its leading exponent, $a_0$ and $a_1$ are nonuniversal amplitudes and $\omega>0$ is a correction-to-scaling exponent. In the fitting procedure, we first allow $\omega$ to vary freely. If the fit is unstable, we then perform fits with a series of fixed $\omega$ values. Even when a stable fit is obtained with free $\omega$, we also explore nearby fixed values of $\omega$ to check the robustness of the fit and identify potentially more accurate results.

To check the stability of the fit, we repeated the fits using only data with $L\geq L_{\min}$ and examined how the estimated parameters vary with $L_{\min}$. In general, our preferred fit corresponds to the smallest $L_{\min}$ for which the goodness of fit is reasonable, and for which further increases of $L_{\min}$ do not reduce chi-square by significantly more than one unit per degree of freedom (DF).

\section{Simulation results}    \label{sec-sr}

\begin{figure}
\centering
\includegraphics[width=\columnwidth]{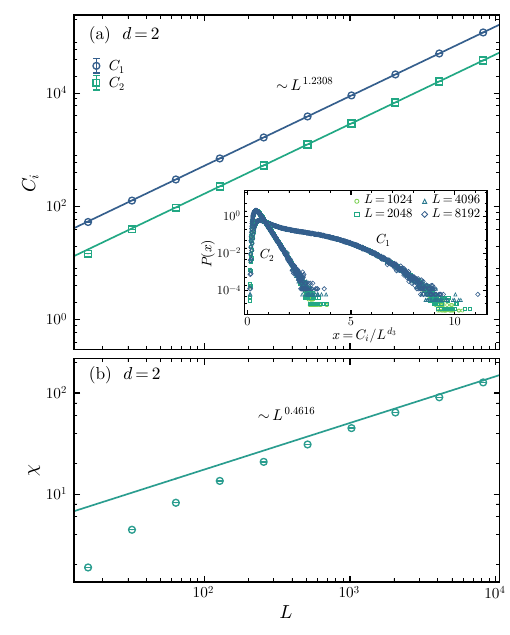}
\caption{FSS of triconnected clusters in two dimensions. (a) The sizes of the largest and second-largest triconnected clusters, $C_1$ and $C_2$, scale as $C_1\sim C_2\sim L^{d_3}$. The solid lines show the scaling with the fit result of $d_3$ in Table~\ref{tab-all}. The inset shows the distributions $P(x)$ of the rescaled cluster sizes $x=C_1/L^{d_3}$, and $x=C_2/L^{d_3}$ for different $L$. (b) The susceptibility-like quantity $\chi$ diverges as $\chi\sim L^{2d_3-d}$. The solid lines show the scaling with the fit result of $d_3$ in Table~\ref{tab-all}.}  \label{f4}
\end{figure}

\subsection{Two dimensions}

We first consider site percolation on the triangular lattice. Figure~\ref{f4}(a) shows the mean sizes of the largest and second-largest triconnected clusters, $C_1$ and $C_2$, as functions of the linear system size $L$. Both exhibit power-law scaling,
\begin{equation}
C_1\sim C_2\sim L^{d_3},
\end{equation}
characterized by a common fractal dimension $d_3$. To estimate $d_3$, we fit the $C_1$ data to the FSS form in Eq.~\eqref{eq-fit}. The results for different fitting ranges and fitting schemes, with the correction exponent $\omega$ either fixed or treated as a free parameter, are summarized in Table~\ref{tab-c1tl}. Taking the consistency among these estimates into account, we obtain $d_3=1.2308(3)$. Fits to the $C_2$ data yield consistent results. The inset of Fig.~\ref{f4}(a) shows the distributions $P(x)$ of the rescaled cluster sizes $x=C_1/L^{d_3}$, and $x=C_2/L^{d_3}$ for different $L$. The good collapse of data from different system sizes reinforces the correctness of our fitting result. We also obtain $d_3=1.231(1)$ for bond percolation on the square lattice, although with lower precision (Table~\ref{tab-d3fit}). The agreement between the two models supports the universality among site and bond percolation.

FSS theory further predicts that the susceptibility-like quantity $\chi$ scales as
\begin{equation}
\chi \sim L^{2d_3-d}.
\end{equation}
Figure~\ref{f4}(b) shows $\chi$ as a function of $L$ for site percolation on the triangular lattice. Although substantial finite-size corrections are present, the data gradually approach the predicted scaling behavior as $L$ increases. The solid line shows the expected power-law scaling, with $d_3=1.2308$ fixed to the value obtained independently from the $C_1$ data. The agreement between the numerical data and this prediction provides an independent consistency check of the scaling description for triconnected clusters.

\begin{table}[b]
\caption{Fit results of $d_3$ for site percolation on triangular lattices. The maximum size is $L_{\max}=8192$. The fractal dimension $d_3$ is obtained by fitting the $C_1$ data to Eq.~\eqref{eq-fit}. The value of $\omega$ without errors indicates a fixed value in the fit.}    \label{tab-c1tl}
\begin{ruledtabular}
\begin{tabular}{llllll}
$L_{\min}$ & \multicolumn{1}{c}{$d_3$} & \multicolumn{1}{c}{$a_0$} & \multicolumn{1}{c}{$\omega$} & \multicolumn{1}{c}{$a_1$} & \multicolumn{1}{c}{$\chi^2/\mathrm{DF}$}  \\
\hline
$32$    & $1.2310(1)$    & $1.799(1)$      & $1.5(2)$  & $-1.7(8)$   & $8.7/5$   \\
$64$    & $1.2301(9)$    & $1.81(2)$       & $0.5(4)$  & $-0.08(4)$  & $3.5/4$   \\
$64$    & $1.2308(1)$    & $1.802(1)$      & $1$       & $-0.30(4)$  & $5.4/5$    \\
$128$   & $1.2307(2)$    & $1.803(2)$      & $1$       & $-0.4(1)$   & $3.9/4$   \\
$64$    & $1.23091(9)$   & $1.800(1)$      & $1.25$    & $-0.71(8)$  & $6.9/5$   \\
$128$   & $1.2307(1)$    & $1.802(2)$      & $1.25$    & $-1.1(3)$   & $4.2/4$   \\
$64$    & $1.23099(8)$   & $1.799(1)$      & $1.5$     & $-1.8(2)$   & $8.5/5$   \\
$128$   & $1.2308(1)$    & $1.801(2)$      & $1.5$     & $-3.3(8)$   & $4.5/4$   \\
$128$   & $1.2309(1)$    & $1.801(1)$      & $1.75$    & $-10(2)$    & $4.8/4$   \\
$256$   & $1.2308(2)$    & $1.802(3)$      & $1.75$    & $-17(12)$   & $4.4/3$
\end{tabular}
\end{ruledtabular}
\end{table}

\begin{table}
\caption{Fit results of $d_3$ for bond percolation on square lattices, cubics, and hypercubics ($d=4$). The maximum sizes are $L_{\max}=8192$, $406$, and $90$, for $d=2$, $3$, and $4$, respectively. The fractal dimension $d_3$ is obtained by fitting the $C_1$ data to Eq.~\eqref{eq-fit}. The value of $\omega$ without errors indicates a fixed value in the fit.}    \label{tab-d3fit}
\begin{ruledtabular}
\begin{tabular}{cllllll}
$d$  & $L_{\min}$ & \multicolumn{1}{c}{$d_3$} & \multicolumn{1}{c}{$a_0$} & \multicolumn{1}{c}{$\omega$} & \multicolumn{1}{c}{$a_1$} & \multicolumn{1}{c}{$\chi^2/\mathrm{DF}$}  \\
\hline
$2$ & $64$    & $1.232(2)$     & $0.70(1)$       & $0.5(2)$  & $0.13(3)$   & $3.6/4$   \\
    & $128$   & $1.231(2)$     & $0.70(1)$       & $0.7(4)$  & $0.2(2)$    & $3.4/3$   \\
    & $128$   & $1.2316(5)$    & $0.695(3)$      & $0.5$     & $0.13(2)$   & $3.6/4$  \\
    & $256$   & $1.2311(9)$    & $0.698(5)$      & $0.5$     & $0.11(3)$   & $3.1/3$  \\
    & $128$   & $1.2304(4)$    & $0.704(2)$      & $0.75$    & $0.27(3)$   & $3.5/4$   \\
    & $256$   & $1.2303(6)$    & $0.704(4)$      & $0.75$    & $0.26(8)$   & $3.4/3$  \\
\hline
$3$ & $25$    & $1.081(1)$     & $0.759(6)$   & $1.6(2)$  & $-7(2)$      & $7.3/3$    \\
    & $64$    & $1.082(2)$     & $0.755(7)$   & $1.25$    & $-1.3(4)$    & $1.0/2$   \\
    & $101$   & $1.081(3)$     & $0.762(1)$   & $1.25$    & $-2(1)$      & $0.6/1$    \\
    & $64$    & $1.083(1)$     & $0.751(6)$   & $1.5$     & $-3.00(79)$  & $1.1/2$     \\
    & $101$   & $1.081(3)$     & $0.76(1)$    & $1.5$     & $-4.9(26)$   & $0.5/1$    \\
    & $40$    & $1.0814(7)$    & $0.757(3)$   & $1.75$    & $-10.9(6)$   & $5.4/3$    \\
    & $64$    & $1.083(1)$     & $0.748(5)$   & $1.75$    & $-7(2)$      & $1.2/2$    \\
    & $101$   & $1.082(2)$     & $0.76(1)$    & $1.75$    & $-13(7)$     & $0.5/1$   \\
\hline
$4$ & $16$    & $0.84(1)$      & $1.3(1)$     & $1.0(1)$  & $-5.4(9)$    & $10.0/2$   \\
    & $22$    & $0.76(6)$      & $2(1)$       & $0.5(3)$  & $-3.6(5)$    & $1.3/1$    \\
    & $22$    & $0.746(2)$     & $2.54(3)$    & $0.4$     & $-3.76(7)$   & $1.3/2$     \\
    & $32$    & $0.749(5)$     & $2.51(7)$    & $0.4$     & $-3.7(2)$    & $1.1/1$  \\
    & $22$    & $0.769(2)$     & $2.11(2)$    & $0.5$     & $-3.54(7)$   & $1.3/2$  \\
    & $32$    & $0.770(5)$     & $2.10(6)$    & $0.5$     & $-3.5(2)$    & $1.3/1$    \\
    & $22$    & $0.788(2)$     & $1.83(2)$    & $0.6$     & $-3.55(7)$   & $1.7/2$    \\
    & $32$    & $0.787(5)$     & $1.84(4)$    & $0.6$     & $-3.6(2)$    & $1.6/1$
\end{tabular}
\end{ruledtabular}
\end{table}

\subsection{Three and four dimensions}

\begin{figure}
\centering
\includegraphics[width=\columnwidth]{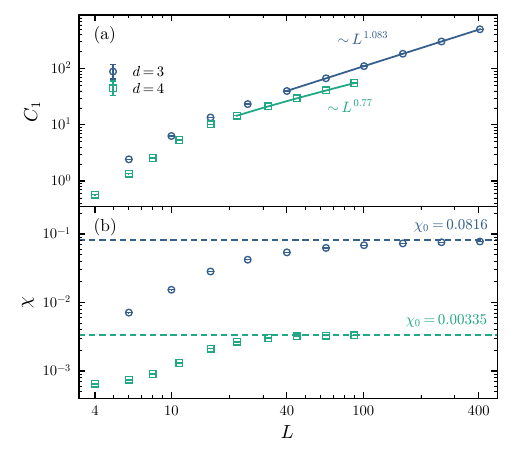}
\caption{FSS of triconnected clusters in dimensions $d=3$ and $4$. (a) The size of the largest triconnected cluster, $C_1$, is plotted as a function of $L$. The solid lines show the scaling with the fit result of $d_3$ in Table~\ref{tab-all}. (b) The susceptibility-like quantity $\chi$ approaches a finite value as $\chi\sim \chi_0 + a L^{2d_3-d}$. The dashed lines represent the fit results of $\chi_0$.}\label{f5}
\end{figure}

We next consider percolation above two dimensions. Figure~\ref{f5}(a) shows the mean size of the largest triconnected cluster, $C_1$, as a function of $L$ for $d=3$ and $4$. The data exhibit an overall growth as the system size increases. However, finite-size corrections become increasingly pronounced as the spatial dimension increases, making it more difficult to obtain precise estimates of $d_3$. In particular, the effective scaling behavior varies substantially over the accessible system sizes, although the data at the largest sizes still show a tendency toward power-law scaling.

In most cases, allowing $\omega$ to vary freely does not yield stable fits to the FSS form in Eq.~\eqref{eq-fit}. We therefore perform fits with several fixed values of $\omega$. The stable fits are summarized in Table~\ref{tab-d3fit}. Since the resulting estimates of $d_3$ vary with the choice of $\omega$, we account for this variation when determining the final uncertainties. Consequently, the estimates of $d_3$ are less precise than those obtained in two dimensions, yielding $d_3=1.083(2)$ and $0.77(2)$ for $d=3$ and $4$.

Despite the larger uncertainties, a clear trend emerges: the estimated fractal dimension $d_3$ decreases systematically with increasing spatial dimension. For comparison, Table~\ref{tab-all} summarizes the fractal dimensions $d_1$, $d_2$, and $d_3$ of ordinary percolation clusters, their backbones, and triconnected clusters, respectively, where $d_1$ and $d_2$ increases with dimensions from $d=2$ to $4$. The systematic decrease of $d_3$ therefore highlights the distinct dimensional dependence of triconnected clusters and further emphasizes the special geometric properties of higher-order connectivity.

Figure~\ref{f5}(b) shows the susceptibility-like quantity $\chi$ for dimensions $d=3$ and $4$. In contrast to $d=2$, $\chi$ does not exhibit a divergence with increasing system size. Instead, it approaches a finite value in each dimension. This behavior is consistent with the scaling prediction for its singular contribution, $\chi\sim L^{2d_3-d}$: when $2d_3-d<0$, the singular contribution decreases with increasing $L$, leaving a finite background contribution that dominates the asymptotic behavior. Thus, the behavior of $\chi$ provides further evidence for the distinct scaling properties of triconnected clusters in three and higher dimensions.

To further examine this prediction, we fit the data to the FSS anaztz
\begin{equation}
\chi(L) = \chi_0 + L^{2d_3-d} (a_0+a_1L^{-\omega}), \label{eq-fchi}
\end{equation}
The resulting estimates of $d_3$ are $1.08(1)$ and $0.75(5)$ for $d=3$ and $4$, respectively, consistent with those obtained independently from the largest-cluster sizes (Table~\ref{tab-all}).

\subsection{Six dimensions and complete graphs}

\begin{figure}
\centering
\includegraphics[width=\columnwidth]{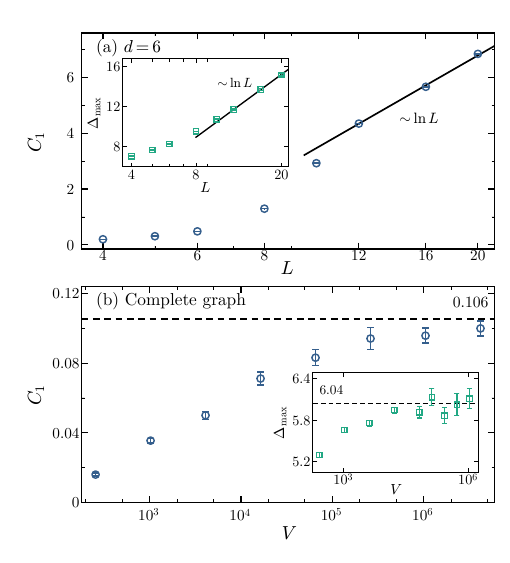}
\caption{(a) The size of the largest triconnected cluster, $C_1$, is plotted as a function of $L$ in dimension $d=6$. The solid lines show the behavior $C_1\sim \ln L$. The inset plots the maximum increment $\Delta_{\max}$ in the bond-addition process against $L$, where the solid represents $\Delta_{\max} \sim \ln L$. (b) The size of the largest triconnected cluster, $C_1$, is plotted as a function of $V$ on complete graphs. The dashed line show the asymptote value of $C_1$. The inset plots the maximum increment $\Delta_{\max}$ in the bond-addition process against $V$, which also appoaches a finite value indicated by the dashed line.}   \label{f6}
\end{figure}

\begin{figure*}
\centering
\includegraphics[width=\textwidth]{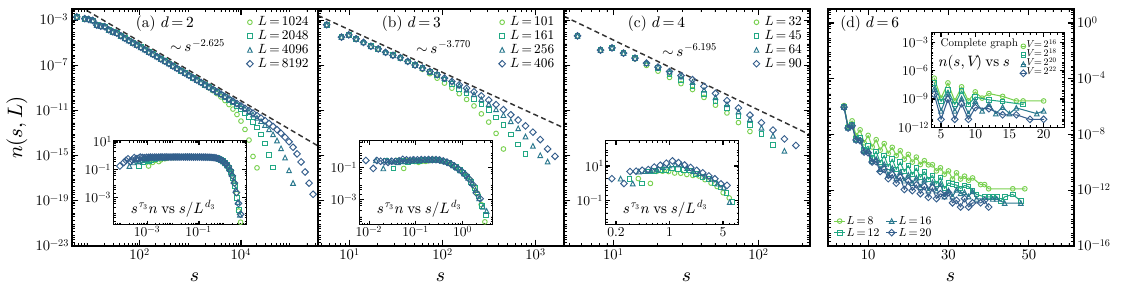}
\caption{Cluster-size distributions $n(s,L)$ of triconnected clusters in dimensions $d=2$, $3$, $4$, and $6$, as well as on complete graphs, for different system sizes. (a–c) Power-law regimes characterized by $n(s,L) \sim s^{-\tau_3}$, indicated by the dashed lines, are observed for $d=2$–$4$. The insets show $s^{\tau_3}n(s,L)$ as a function of $s/L^{d_3}$, where $d_3$ is taken from the estimates in Table~\ref{tab-all}. (d) In six dimensions, where $d_3=0$, the distribution dose not exhibit a scaling regime, and overall, it decreases with the sytem size increases. The inset shows the results for complete graphs, where only small triconnected clusters are observed, and also decrease with the sytem size increases.}  \label{f7}
\end{figure*}

We now examine the behavior of triconnected clusters in six dimensions and on complete graphs. Figure~\ref{f6}(a) shows the size of the largest triconnected cluster, $C_1$, as a function of $L$ in a semi-log plot. The data demonstrate a straight line for increasing system size, suggesting a logarithmic growth, $C_1\sim \ln L$. Thus, although the largest triconnected cluster continues to grow with system size, its growth is slower than any positive power of $L$. In terms of the conventional power-law definition $C_1\sim L^{d_3}$, this corresponds to a vanishing fractal dimension, $d_3=0$.

For complete graphs, in a significant fraction of realizations, there are no triconnected clusters at criticality, since a triconnected cluster must contain at least $4$ sites. Averaging over all samples thus yields a very small $C_1$, less than $1$, as shown in Fig.~\ref{f6}(b). In this case, $C_1$ tends to a finite value as the system size increases, indicating that the largest triconnected cluster does not grow asymptotically. Its fractal dimension is therefore also zero. The comparison between six-dimensional lattices and complete graphs reveals two distinct forms of sub-power-law behavior: logarithmic growth in six dimensions and saturation to a finite size on complete graphs. 

To further investigate whether the percolation transition involves a discontinuous change in the triconnected-cluster order parameter when $d\geq6$, we examine the largest increment of $C_1$ during a bond-addition process~\cite{Li2023,Lu2026}. Specifically, we add the bonds one at a time $t$ and monitor $C_1(t)$. At each step, we record the increment $\Delta=C_1(t)-C_1(t-1)$, and determine the maximum value $\Delta_{\max}$, over the entire process. The insets of Fig.~\ref{f6} show the mean unnormalized maximum increment $\Delta_{\max}$, against $L$ in six dimensions and against $V$ on complete graphs. In six dimensions, the maximum increment increases approximately as $\ln L$, whereas on complete graphs it approches a finite value. These results indicate that the triconnected-cluster order parameter, $\sim C_1/V$, does not exhibit a finite jump in either case, supporting a continuous rather than discontinuous onset.

\subsection{Cluster size distributions across dimensions}

Finally, we examine the size distribution of triconnected clusters, $n(s,L)$, where $n(s,L)$ denotes the number of triconnected clusters of size $s$ per lattice site. Figure~\ref{f7} shows the distributions for dimensions $d=2$, $3$, $4$, and $6$, as well as for complete graphs. For $d=2$-$4$, the distributions exhibit power-law behaviors over intermediate ranges of cluster sizes, followed by a cutoff at large $s$. The data are consistent with the scaling form
\begin{equation}
n(s,L) = s^{-\tau_3} \tilde{n}(s/L^{d_3}),    \label{eq-ns}
\end{equation}
where the Fisher exponent is related to the fractal dimension through
\begin{equation}
\tau_3=1+\frac{d}{d_3}.
\end{equation}
The Fisher exponents $\tau_3$, estimated using the values of $d_3$ in Table~\ref{tab-all} and the scaling relation $\tau_3=1+d/d_3$, are consistent with the power-law regimes observed in $n(s,L)$ in Fig.~\ref{f7}. Furthermore, as shown in the insets of Figs.~\ref{f7}(a–c), plotting $s^{\tau_3}n(s,L)$ against $s/L^{d_3}$ leads to a good data collapse for different system sizes, with clear plateaus corresponding to the scaling regimes. These results provide further evidence for well-defined fractal scaling of triconnected clusters in low dimensions.

An interesting feature is that the power-law regime becomes progressively narrower as the spatial dimension increases. This trend is consistent with the decreasing fractal dimension $d_3$: as $d_3$ decreases, the cutoff size $\sim L^{d_3}$ of $n(s,L)$ becomes small, the range of cluster sizes exhibiting clear scale-invariant behavior thus becomes increasingly restricted.

As shown in Fig.~\ref{f7}(d) for $d=6$, no clear power-law regime can be identified in the cluster-size distribution. We therefore conjecture that the distribution has an exponential tail,
\begin{equation}
n(s,L)\sim e^{-as},
\end{equation}
where $a^{-1}$ characterizes the cutoff. This form is consistent with the observed logarithmic growth of $C_1$: according to extreme-value statistics, if the system contains an extensive number of approximately independent candidate clusters, the expected number of clusters with sizes of order $C_1$ is $\mathcal{O}(1)$, giving
\begin{equation}
L^d e^{-aC_1}\sim O(1),
\end{equation}
and hence $C_1\sim\ln L$.

For complete graphs, the cluster-size distribution shown in the inset of Fig.~\ref{f7}(d) is also strongly suppressed, with the observed cluster sizes confined to a small range. As shown in Fig.~\ref{f6}(b), the largest triconnected cluster approaches a finite value as the system size increases. The distribution is therefore expected to exhibit a cutoff at a finite cluster size in the thermodynamic limit, in contrast to the logarithmic growth of the largest cluster in six dimensions.

This comparison motivates the conjecture that the exponential tail and the resulting logarithmic growth of the largest triconnected cluster persist in finite-dimensional lattices above the upper critical dimension $d_u=6$, whereas the largest cluster remains finite in the complete-graph limit. This distinction can be understood in terms of the number of biconnected clusters. Previous studies have shown that their number scales extensively with the system volume $V$ in finite-dimensional lattices, whereas it grows only logarithmically with $V$ on complete graphs~\cite{Huang2018}. Thus, finite-dimensional lattices provide an extensive number of opportunities for forming large triconnected structures, whereas the much slower growth in the number of biconnected clusters on complete graphs suppresses their formation. This picture is consistent with the different behaviors of $C_1$ observed in the two cases, although the underlying mechanism requires further investigation.

\section{Discussion}  \label{sec-dis}

In this work, we have investigated the critical geometry of triconnected clusters in percolation, introducing an efficient numerical approach based on the SPQR decomposition of graphs. By applying this approach to percolation in different dimensions, we have systematically characterized the fractal properties of triconnected clusters. Our results reveal a distinct dimensional dependence of their geometry, with the fractal dimension $d_3$ decreasing as the spatial dimension increases. We have also examined the associated cluster-size distributions and susceptibility-like quantities, providing complementary tests of the FSS of triconnected clusters.

The decrease of $d_3$ with increasing dimension is particularly interesting. It contrasts with the behavior commonly observed for ordinary percolation clusters and their backbones, whose fractal dimensions generally increase from $d=2$ to $4$. A possible qualitative explanation is related to the increasingly tree-like geometry of critical clusters at high dimensions. Ordinary connectivity requires only a path between sites, whereas triconnectivity requires multiple independent paths, which become relatively scarce as critical clusters approach a tree-like geometry. Consequently, triconnected structures occupy a progressively smaller fraction of the critical clusters as the dimension increases. We point out that this effect due to high-order connectivities can even lead to a nonmonotonic variation of fractal dimensions as the dimension increases.~\cite{Wang2026}.

This trend becomes especially pronounced in six dimensions and on complete graphs. In six dimensions, the largest triconnected cluster grows logarithmically with system size, corresponding to a vanishing fractal dimension. On complete graphs, its size instead approaches a finite limit. Despite these different asymptotic behaviors, a continuous onset of triconnectivity is still observed.

Several directions merit further investigation. First, it would be interesting to extend the present study to four-connected, and even higher-order structures, and to examine whether their fractal dimensions vanish at lower spatial dimensions. In two dimensions, such studies would also provide numerical benchmarks for arm exponents in percolation, which may help advance their understanding within the SLE framework. Second, it would be valuable to investigate higher-connectivity structures in the FK representation of the $q$-state Potts model, whose $q\to1$ limit corresponds to percolation. Studying triconnected components in this representation may provide a complementary perspective on the geometric organization of critical systems.

\section*{Acknowledgements}

M.L. and Q.W. were supported by the National Natural Science Foundation of China under Grant No.~12675037. Y.D. was supported by the National Natural Science Foundation of China under Grant No.~12275263, and by the Quantum Science and Technology National Science and Technology Major Project under Grant No.~2021ZD0301900.

\textit{Note added}. After we had completed our numerical analysis, Gefei Cai brought his recent independent work~\cite{Cai2026} to our attention, in which the exact value of the monochromatic three-arm exponent in two-dimensional percolation, $\xi_3=0.76928375\ldots$, is derived analytically. Thus, our independent studies constitute a double-blind verification of this result, with excellent agreement between the analytical and numerical estimates through the relation $d_3=d-\xi_3$.

\bibliography{references}

\end{document}